\documentclass[preprints,article,accept,moreauthors,pdftex]{Definitions/mdpi}

\usepackage{color}
\usepackage[T1]{fontenc}
\usepackage[utf8]{inputenc}
\usepackage{alphabeta}
\usepackage{graphicx}
\usepackage{svg}
\usepackage{amsmath}
\usepackage{graphicx}
\usepackage{dcolumn}
\usepackage{bm}
\firstpage{1} 
\pubvolume{1}
\issuenum{1}
\articlenumber{0}
\pubyear{2021}
\copyrightyear{2020}
\datereceived{} 
\dateaccepted{} 
\datepublished{} 
\hreflink{https://doi.org/} 
\newtheorem{exampleC}{Example}
\newtheorem{mydef}{Definition}
\newtheorem{lemmaC}{Lemma}

\Title{A Language for Modeling And Optimizing Experimental Biological Protocols}

\TitleCitation{A Language for Modeling And Optimizing Experimental Biological Protocols}

\Author{Luca Cardelli $^{1}$, Marta Kwiatkowska $^{1}$ and Luca Laurenti $^{1,\dagger}$}

\AuthorNames{Luca Cardelli, Marta Kwiatkowska and Luca Laurenti}

\AuthorCitation{Cardelli, L.; Kwiatkowska, M.; Laurenti, L.}

\address{%
$^{1}$ \quad Affiliation: Department of Computer Science, University of Oxford\\
}

\firstnote{Current address: Delft Center for Systems and Control (DCSC), TU Delft} 

\abstract{Automation is becoming ubiquitous in all laboratory activities, leading towards precisely defined and codified laboratory protocols. However, the integration between laboratory protocols and mathematical models is still lacking. Models describe physical processes, while protocols define the steps carried out during an experiment: neither cover the domain of the other, although they both attempt to characterize the same phenomena. We should ideally start from an integrated description of both the model and the steps carried out to test it, to concurrently analyze uncertainties in model parameters, equipment tolerances, and data collection. To this end, we present a language to model and optimize experimental biochemical protocols that facilitates such an integrated description, and that can be combined with experimental data. {  We provide a probabilistic semantics for our language in terms of Gaussian processes (GPs) based on the Linear Noise Approximation (LNA) that  formally characterizes the uncertainties in the data collection, the underlying model, and the protocol operations. }On a set of case studies we illustrate how the resulting framework allows for automated analysis and optimization of experimental protocols, including Gibson assembly protocols.}

\keyword{Chemical Reaction Networks, Gaussian Processes, Biological Protocols} 

\begin{document}

\section{Introduction}

\textbf{State of the art in lab automation.}
Automation is becoming ubiquitous in all laboratory activities: protocols are run under reproducible and auditable software control, data are collected by high-throughput machinery, experiments are automatically analyzed, and further experiments are selected to maximize knowledge acquisition. However, while progress is being made towards the routine integration of sophisticated end-to-end laboratory workflows and towards the remote access to laboratory facilities and procedures \cite{murphy2018synthesizing,ananthanarayanan2010biocoder,cardelli2017syntax,ang2013tuning,abate2018experimental}, the integration between laboratory protocols and mathematical models is  still lacking. Models describe physical processes, either mechanistically or by inference from data, while protocols define the steps carried out during an experiment in order to obtain experimental data. Neither models nor protocols cover the domain of the other, although they both attempt to characterize the same phenomena. As a consequence, it is often hard to attribute causes of experimental failures: whether an experiment failed because of a misconception in the model, or because of a misstep in the protocol. To confront this problem we need an approach that integrates and accounts for all the components, theoretical and practical, of a laboratory experiment. We should ideally start from an integrated description from which we can extract both the model of a phenomenon, for possibly automated mathematical analysis, and the steps carried out to test it, for automated execution by lab equipment. This is essential to enable automated model synthesis and falsification by concurrently taking into account uncertainties in model parameters, equipment tolerances, and data collection. 

\begin{figure}
    \centering
    \includegraphics[width=0.70\textwidth]{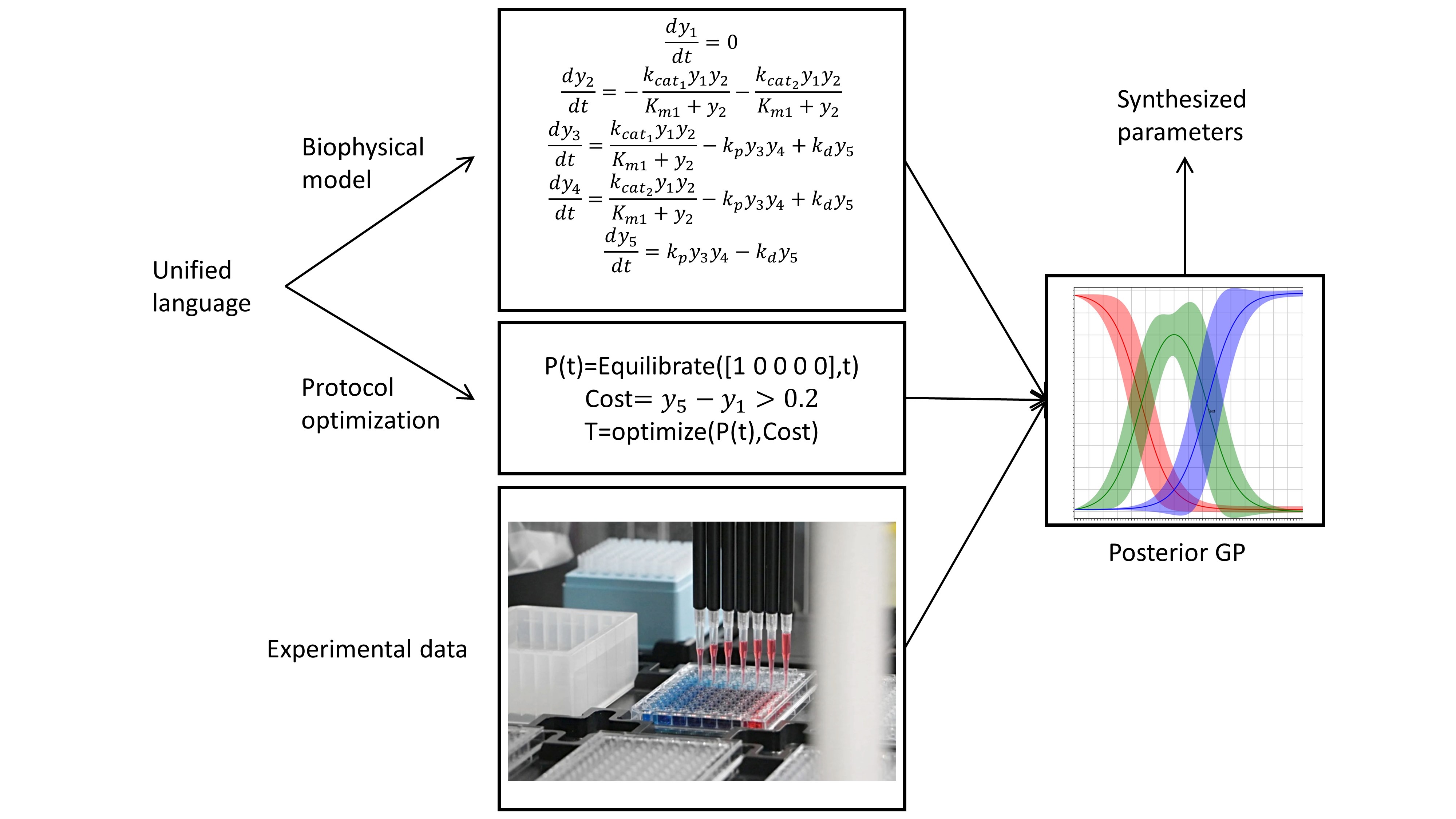}
    \caption{Processing a unified description for an experimental protocol.  A program integrates a biophysical model of the underlying molecular system with the steps of the protocol. In this case the protocol comprises a single instruction, which lets a sample equilibrate for $t$ seconds, where $t$ is a parameter.
    The initial concentration of the sample is $1$ for the first species and $0$ for all the others. The value of $t$ is selected as the one the maximizes a cost function, in this case the difference between $y_5$ and $y_1$ after the execution of the protocol. The optimization is performed on a Gaussian process given by the semantics of the program (biophysical model and protocol) integrated with experimental data.}
    \label{fig:Gibson}
\end{figure}

\textbf{Our approach.}
We present a language to model and optimize experimental biochemical protocols that provides such an integrated description of the protocol and of the underlying molecular process, and that can be combined with experimental data. From this integrated representation, both the model of a phenomenon and the steps carried out to test it can be separately extracted. 
Our approach is illustrated in Figure \ref{fig:Gibson}. 
We provide a probabilistic semantics for our language in terms of a Gaussian process (GP) \cite{rasmussen2003gaussian}, which can be used to characterize uncertainties. Such a semantics arises from a Linear Noise Approximation (LNA) \cite{Kampen1992b,cardelli2016stochastic} of the dynamics of the underlying biochemical system and of the protocol operations, and it corresponds to a Gaussian noise assumption. We show that in a Bayesian framework the resulting semantics can be combined in a principled way with experimental data to build a posterior process  that integrates  our prior knowledge with new data. We demonstrate that the Gaussian nature of the resulting process allows one to efficiently and automatically optimize the parameters of an experimental protocol in order to maximize the performances of the experiment.  On a series of case studies, including a Gibson Assembly protocol \cite{gibson2009enzymatic} and a Split and MIx protocol, we highlight the usefulness of our approach and how it can have an impact on scientific discovery.

\textbf{Related work and novelty.} 
Several factors contribute to the growing need for a formalization of experimental protocols in biology. First, better record keeping of experimental operations is recognized as a step towards tackling the ‘reproducibility crisis’ in biology \cite{begley2012raise}. Second, the emergence of ‘cloud labs’ creates a need for precise, machine-readable descriptions of the experimental steps to be executed. To address these needs, frameworks allowing protocols to be recorded, shared, and reproduced locally or in a remote lab have been proposed. These frameworks introduce different programming languages for experimental protocols, including BioScript \cite{ott2018bioscript,baker20161}, BioCoder \cite{ananthanarayanan2010biocoder}, Autoprotocol \cite{autoprot}, and Antha \cite{antha}. These languages provide expressive, high-level protocol descriptions but consider each experimental sample as a labelled ‘black-box’. This makes challenging the study a protocol together with the biochemical systems it manipulates in a common framework.
In contrast, we consider a simpler set of protocol operations but we capture the details of experimental samples, enabling us to track properties of chemical solutions (concentrations and temperatures) as the chemicals wherein react during the execution of a protocol. This allows us to formalize and verify requirements for the correct execution of a protocol and to optimize various protocol or system parameters to satisfy these specifications. 

Our language derives from our previous conference paper \cite{abate2018experimental} by introducing stochasticity in chemical evolution, and by providing a Gaussian  semantics, particularly for protocol operations, for the effective analysis of the combined evolution of stochastic chemical kinetics and protocols and experimental data. The language semantics has been already implemented (but not previously formally presented) in a chemical/protocol simulator \cite{cardelli2020kaemika}. We expect that this style of semantics could be adapted to other protocol languages, along the principles we illustrate, to provide a foundation for the analysis of complex stochastic protocols in other settings.

\textbf{Paper outline.}
In what follows, we first introduce the syntax and semantics of our language. We then show how we can perform  
optimization of the resulting Gaussian process integrated with data. Finally, we illustrate the usefulness of our approach on several case studies, highlighting the potential impact of our work {on scientific discovery}.

\section{Materials and Methods}
In this section we introduce the syntax of the language we propose for modelling experimental protocols. A formal semantics of the language, based on denotational semantics \cite{scott1971toward}, is then discussed. The physical process underlying a biological experimental protocol is modeled  as a Chemical Reaction Systems (CRS). As a consequence, before introducing the language for experimental protocols, we first formally introduce Chemical Reaction Networks (CRNs) and Chemical Reaction Systems (CRSs).

\subsection{Chemical Reaction Network (CRN)}

 Chemical Reaction Networks (CRNs) is a standard language for modelling biomolecular interactions \cite{Cardelli2010}.
\begin{mydef}(CRN)
\label{CRNDef}
A \emph{chemical reaction network} $\mathcal{C}=(\mathcal{A},\mathcal{R}) \in CRN =  \Lambda\times\mathcal{T}$ is a pair of a finite set $\mathcal{A} \subseteq \Lambda$ of \emph{chemical species}, of size $|\mathcal{A}|$, and a finite set $\mathcal{R}\subseteq \mathcal{T}$ of \emph{chemical reactions}. 
A \emph{reaction} $\tau \in \mathcal{R}$ is a triple $\tau=(r_{\tau},p_{\tau},k_{\tau})$, where $r_{\tau} \in  \mathbb{N}^{|\Lambda|}$ is the \emph{source complex}, $p_{\tau} \in  \mathbb{N}^{|\Lambda|}$ is the \emph{product complex} and $k_{\tau} \in \mathbb{R}_{>0} $ is the coefficient associated with the rate of the reaction.
The quantities $r_{\tau}$ and $p_{\tau}$  are the stoichiometry of reactants and products. The \emph{stoichiometric vector} associated to $\tau$ is defined by $\upsilon_{\tau}=p_{\tau} - r_{\tau}$.
Given a reaction $\tau_i=(  [1,0,1],[0,2,0],k_i )$ we refer to it visually as $\tau_i : \lambda_1 + \lambda_3 \, \rightarrow^{k_i}  \,    2\lambda_2 $.
\end{mydef}

\begin{mydef}(CRN State)
\label{CRNStateDef}
Let $\mathcal{C}=(\mathcal{A},\mathcal{R})\in CRN$. A \emph{state} of $\mathcal{C}$, of the form $(\mu, \Sigma, V, T)$ $\in S=(\mathbb{R}_{\geq 0}^{|\mathcal{A}|}\times\mathbb{R}^{|\mathcal{A}|\times|\mathcal{A}|}\times\mathbb{R}_{\geq 0}\times\mathbb{R}_{\geq 0})$, consists of a concentration vector $\mu$ (moles per liter), a covariance matrix $\Sigma$, a volume (liters), and a temperature (degrees Celsius). A CRN together with a (possibly initial) state is called a \emph{Chemical Reaction System} (CRS) having the form $(\mathcal{A},\mathcal{R}),(\mu, \Sigma, V, T)$.
\end{mydef}
\noindent
A Gaussian process is a collection of random variables, such that every finite linear combination of them is normally distributed. Given a state of a CRN its time evolution from that state can be described by a Gaussian process indexed by time whose mean and variance are given by a Linear Noise Approximation (LNA) of the Chemical Master Equation (CME) \cite{bortolussi2019central,Kampen1992b} and is formally defined in the following definitions of CRN Flux and CRN Time Evolution.

\begin{mydef}(CRN Flux)
\label{CRNFluxDef}
Let $(\mathcal{A},\mathcal{R})$ be a CRN. Let $F(V,T)\in \mathbb{R}^{|\mathcal{A}|}_{\geq 0}\to\mathbb{R}^{|\mathcal{A}|}$ be the flux of the CRN at volume $V\in \mathbb{R}_{\geq 0}$ and temperature $T\in \mathbb{R}_{\geq 0}$. For a concentration vector $\mu\in \mathbb{R}^{|\mathcal{A}|}_{\geq 0}$ we assume $ F(V,T)(\mu) = \sum_{\tau\in\mathcal{R}}\upsilon_{\tau}\alpha_{\tau}(V,T,\mu)$, with stoichiometric vector $\upsilon_{\tau}$ and rate function $\alpha_{\tau}$. We call ${J}_F$ the Jacobian of $F(V,T)$, and $J^\top_F$ its transpose. Further, define $W(V,T)(\mu) = \sum_{\tau\in\mathcal{R}}\upsilon_{\tau}\upsilon_{\tau}^{\top}\alpha_{\tau}(V,T,\mu)$ to be the diffusion term.
\end{mydef}

\noindent
We should remark that, if one assumes mass action kinetics, then for a reaction $\tau$ it holds that $\alpha_{\tau}(V,T,\mu) = k'_{\tau}(k_{\tau},V,T)\mu^{r_{\tau}}$ where $\mu^{s} = \prod_{i=1}^{|\Lambda|} \mu_i^{s_i}$ is the product of the reagent concentrations, and $k'_{\tau}$ encodes any dependency of the reaction rate $k_{\tau}$ on volume and temperature. Then the Jacobian evaluated at $\mu$ is given by $J_{F_{ik}}(\mu) =  \frac{\partial F(V,T)(\mathbf{x})_{i}}{\partial \mathbf{x}_k}\biggr\rvert_{\mathbf{x}=\mu} = \sum_{\tau\in\mathcal{R}}\upsilon_{\tau_i}r_{\tau_k}\frac{\alpha_{\tau}(V,T,\mu)}{\mu_{k}}$ for $i,k \in |\mathcal{A}|$ \cite{cardelli2016stochastic}.

In the following, we use $\mu,\Sigma$ for concentration mean vectors and covariance matrices respectively, and $\bm{\mu}, \bm{\Sigma}$ (boldface) for the corresponding functions of time providing the evolution of means and covariances.

\begin{mydef}(CRN Time Evolution)
\label{CRSGaussDef} 
Given a CRS $(\mathcal{A},\mathcal{R}),(\mu,\Sigma, V, T)$, its evolution at time $t < H$ (where $H\in \mathbb{R}_{\geq 0}\cup\{\infty \}$ is a time horizon) is the state $(\bm{\mu}_\mu(t),\bm{\Sigma}_{\mu,\Sigma}(t),V,T)$ obtained by integrating its flux up to time $t$, where:
\begin{align}
\label{Eqn:Mean}
&\bm{\mu}_\mu(t)=\mu + \int_{0}^{t} F(V,T)(\bm{\mu}_\mu(s))ds \\
\label{Eqn:Variance}
&  \bm{\Sigma}_{\mu,\Sigma}(t)=\Sigma+\int_{0}^t
		J_F(\bm{\mu}_\mu(s))\bm{\Sigma}_{\mu,\Sigma}(s) + \bm{\Sigma}_{\mu,\Sigma}(s)J^\top_F(\bm{\mu}_\mu(s))+W(V,T)(\bm{\mu}_\mu(s))ds,
\end{align}  
with $\bm{\mu}_\mu(0)=\mu$ and $\bm{\Sigma}_{\mu,\Sigma}(0)=\Sigma$.
If, for such an $H$, $\bm{\mu}$ or $\bm{\Sigma}$ are not unique, then we say that the evolution is ill-posed.
Otherwise, $\bm{\mu}_\mu(t)$ and $\bm{\Sigma}_{\mu,\Sigma}(t)$ define a Gaussian process with that mean  and covariance matrix  for $t < H$.
\end{mydef}
An ill-posed problem may result from technically expressible but anomalous CRS kinetics that does not reflect physical phenomena, such as deterministic trajectories that are not uniquely determined by initial conditions, or trajectories that reach infinite concentrations in finite time.

\begin{exampleC}
\label{RunningExample}
Consider the CRN $\mathcal{C}=(\{a,b,c\},\mathcal{R})$, where $\mathcal{R}=\{ ([1,1,0],[0,2,0],0.2),([0,1,1],$ $[0,0,2],0.2) \}$.
Equivalently, $\mathcal{R}$ can be expressed as
$$a + b \rightarrow^{0.2} b + b \quad \quad \quad 
b + c \rightarrow^{0.2} c + c. $$
Then, we have 
\begin{align*}
    & F(a,b,c)=0.2\begin{bmatrix}
- a\cdot b \\
 a\cdot b -  b\cdot c \\
 b\cdot c \\
\end{bmatrix} \\
&W(a,b,c)=0.2\begin{bmatrix}
 a\cdot b & - a\cdot b & 0 \\
- a\cdot b &  a\cdot b +  b\cdot c  & - b\cdot c \\
0 &  - b\cdot c &  b\cdot c
\end{bmatrix} \\
& J_F(a,b,c)=0.2\begin{bmatrix}
- b & - a & 0\\
 b &  a -  c &  -  b  \\
0 &  c   &  b  \\
\end{bmatrix} 
\end{align*}
Consider an initial condition of $\mu_a=0.1,$ $\mu_b=\mu_c=0.001$, then
the time evolution of mean and variance for the GP defined in Definition \ref{CRSGaussDef} are reported in Figure \ref{fig:Running}.
\begin{figure}
\centering
\includegraphics[width=0.5\linewidth]{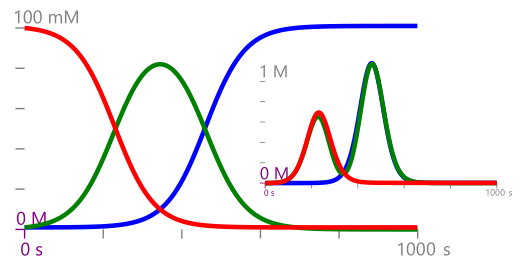}
\caption{
Evolution of $\mu_a$ (red), $\mu_b$ (green), $\mu_c$ (blue) for the CRN reported in Example \ref{RunningExample}. Inset: the respective variances.}
\label{fig:Running}
\end{figure}
\end{exampleC}


\subsection{A Language for Experimental Biological Protocols}
In what follows we introduce the syntax of our language for experimental biological protocols.
\begin{mydef}{(Syntax of a Protocol)}\label{LangSynt2} Given a set of sample variables $x \in Var$ and a set of parameter variables $z \in Par$, the syntax of a protocol $P \in Prot$ for a given fixed CRN $\mathcal{C}=(\mathcal{A},\mathcal{R})$ is 
\begin{align*}
P=\quad \quad \quad\quad \quad \quad\quad\quad \quad \quad \quad \quad \quad x   \quad &\text{(a sample variable)}\\
 (p_1...p_{|\mathcal{A}|},r_V,r_T)\quad  & \text{(a sample with initial concentrations, volume, temperature)}\\
 let\, x=P_1 \, in \, P_2 \quad & \text{(introduce a local sample variable x)} \\
 Mix(P_1,P_2) \quad & \text{(mix samples)}\\
let \, x_1,x_2=Split(P_1,p)\, in\, P_2\quad & \text{(split a sample $P_1$ by a proportion p in (0..1))}\\
 Equilibrate(P,p) \quad & \text{(equilibrate a sample for $p$ seconds)}\\
 Dispose(P) \quad & \text{(discard sample)}\\
 \\
 p=\quad \quad \quad\quad \quad \quad\quad\quad \quad \quad \quad \quad \quad z   \quad &\text{(a parameter variable)}\\
  r \quad & \text{(a literal non-negative real number)}\\
\end{align*}
Moreover, let-bound variables $x$, $x_1$, $x_2$ must occur (as free variables) exactly once in $P_2$.
\end{mydef}
\noindent
A protocol $P$ manipulates \emph{samples} (which are CRN states as in Definition \ref{CRNStateDef}) through a set of operations, and finally yields a sample as the result. This syntax allows one to create and manipulate new samples using Mix (put together different samples), Split (separate samples) and Dispose (discard samples) operations.
Note that the CRN is common to all samples, but different samples may have different initial conditions and hence different active reactions. The single-occurrence (linearity) restriction of sample variables implies that a sample cannot be duplicated or forgotten.







\subsection{Gaussian Semantics for Protocols}
There are two possible approaches to a Gaussian semantics of protocols, that is, to a semantics characterized by keeping track of mean and variance of sample concentrations. They differ in the interpretation of the source of noise that produces the variances. We discuss the two options, and then choose one of the two based on both semantic simplicity and relevance to the well-mixed-solution assumption of chemical kinetics.

The first approach we call \emph{extrinsic noise}. The protocol operations are deterministic, and the the evolution of each sample is also deterministic according to the Rate Equation (that is, Definition \ref{CRSGaussDef}(1)). Here we imagine running the protocol multiple times over a distribution of initial conditions (i.e., the noise is given \emph{extrinsically} to the evolution). The outcome is a final distribution that is determined uniquely by the initial distribution and by the deterministic evolution of each run. For example, the sample-split operation here is interpreted as follows. In each run we have a sample whose concentration in general differs from its mean concentration over all runs. The two parts of a split are assigned identical concentration, which again may differ from the mean concentration. Hence, over all runs, the two parts of the split have identical variance, and their distributions are perfectly correlated, having the same concentration in each run. The subsequent time evolution of the two parts of the split is deterministic and hence identical (for, say, a $50\%$ split). In summary, in the extrinsic noise approach, the variance on the trajectories is due to the distribution of deterministic trajectories for the different initial conditions. 

The second approach we call \emph{intrinsic noise}. The protocol operations are again deterministic, but the evolution in each separate sample is driven by the Chemical Master Equation (i.e., the noise is \emph{intrinsic} to the evolution). Here we imagine running the protocol many times on the same initial conditions, and the variance of each sample is due ultimately to random thermal noise in each run. This model assumes a Markovian evolution of the underlying stochastic system, implying that no two events may happen at the same time (even in separate samples). Also, as usual, we assume that chemical solutions are well-mixed: the probability of a reaction (a molecular collision) is independent of the initial position of the molecules in the solution. Moreover, the well-mixture of solutions is driven by uncorrelated thermal noise in separate samples. Given two initially identical but separate samples, the first chemical reaction in each sample (the first ``fruitful'' collision that follows a usually large number of mixing collisions) is determined by uncorrelated random processes, and their first reaction cannot happen at exactly the same time. Therefore, in contrast to the extrinsic noise approach, a sample-split operation results in two samples that are completely uncorrelated, at least on the time scale of the first chemical reaction in each sample. In summary, in the intrinsic noise approach, the variance on the trajectories is due to the distribution of stochastic trajectories for identical initial conditions.

In both approaches, the variance computations for mix and for split are the same: \emph{mix} uses the squared-coefficient law to determine the variance of two mixed samples, while \emph{split} does not change the variance of the sample being split. The only practical difference is that in the intrinsic noise interpretation we do not need to keep track of the correlation between different samples: we take it to be always zero in view of the well-mixed-solution assumption. { As a consequence, each sample needs to keep track of its internal correlations represented as a local $\Sigma$ matrix of size $|\mathcal{A}|\times|\mathcal{A}|$, but not of its correlations with other samples, which would require a larger global matrix of potentially unbounded size. 
Therefore, the intrinsic noise interpretation results in a vast simplification of the semantics (Definition \ref{GaussSem}) that does not require keeping track of correlations of concentrations across separate samples. } In the rest of the paper we consider only the intrinsic noise semantics.

\begin{mydef}{(Gaussian Semantics of a Protocol - Intrinsic Noise)}\label{GaussSem}

The intrinsic-noise Gaussian semantics $[\![P]\!]^{\rho} \in Prot\times Env\to S$ of a protocol $P \in Prot$ for a CRN $\mathcal{C}=(\mathcal{A},\mathcal{R})$, under environment $\rho \in Env = (Var \cup Par) \to S$, for a fixed horizon $H$ with no ill-posed time evolutions, denotes the final CRN state $(\mu, \Sigma, V, T) \in S$ (Definition \ref{CRNStateDef}) of the protocol and is defined inductively as follows:
\begin{align*}
    &[\![x]\!]^{\rho}=\rho(x)\\
     &[\![(p_1...p_{|\mathcal{A}|},r_V,r_T)]\!]^{\rho}=([\![p_1]\!]^{\rho}...[\![p_{|\mathcal{A}|}]\!]^{\rho},0^{|\mathcal{A}|\times|\mathcal{A}|},r_V,r_T)\\
     &[\![let \, x=P_1\, in\, P_2]\!]^{\rho}=[\![P_2]\!]^{\rho_1}\\
      &\quad where\quad \,{\rho_1}=\rho\{x \leftarrow [\![P_1]\!]^{\rho}\} \\
     &[\![ Mix(P_1,P_2)]\!]^{\rho}= (\frac{V_1\mu_1 +V_2\mu_2}{V_1+V_2},\frac{V_1^2\Sigma_1 +V_2^2\Sigma_2  }{(V_1+V_2)^2},V_1+V_2,\frac{V_1T_1+V_2T_2}{V_1+V_2} )\\
     &\quad where\quad \, (\mu_1,\Sigma_1,V_1,T_1)= [\![P_1]\!]^{\rho}\,\quad and\quad \, (\mu_2,\Sigma_2,V_2,T_2)= [\![P_2]\!]^{\rho}\\
      &[\![let \, x,y=Split(P_1,p)\, in\, P_2]\!]^{\rho}=[\![P_2]\!]^{\rho_1}\\
      &\quad where\quad r = [\![p]\!]^{\rho}, \quad 0 < r < 1 \quad and\quad \,(\mu,\Sigma,V,T)= [\![P_1]\!]^{\rho}\\ 
      & \quad and\quad \,{\rho_1}=\rho\{x \leftarrow (\mu,\Sigma,rV,T),y \leftarrow (\mu,\Sigma,(1-r)V,T)\}\\
       &[\![ Equilibrate(P,p)]\!]^{\rho}= (\bm{\mu}_\mu(t),\bm{\Sigma}_{\mu,\Sigma}(t),V,T)  \\
       &\quad where\quad t = [\![p]\!]^{\rho} \quad and \quad (\mu,\Sigma,V,T)=[\![P]\!]^{\rho}\\
      &[\![Dispose(P)]\!]^{\rho}=(0^{|\mathcal{A}|},0^{|\mathcal{A}|\times|\mathcal{A}|},0,0)
\end{align*} 
together with $[\![p]\!]^{\rho}$ defined as:
\begin{align*}
    &[\![z]\!]^{\rho}=\rho(z)\\
    &[\![r]\!]^{\rho}=r
\end{align*} 

The semantics of $Equilibrate$ derives from Definition \ref{CRSGaussDef}. The substitution notation $\rho\{x \leftarrow v\}$ represents a function that is identical to $\rho$ except that at $x$ it yields $v$; this may be extended to the case where $x$ is a vector of distinct variables and $v$ is a vector of equal size. 
The variables introduced by $let$ are used linearly (i.e., must occur exactly once in their scope), implying that each sample is consumed whenever used.

We say that the components $(\mu,\Sigma)\in \mathbb{R}_{\geq 0}^{|\mathcal{A}|}\times\mathbb{R}^{|\mathcal{A}|\times|\mathcal{A}|}$ of a CRN state $(\mu,\Sigma,V,T)\in S$ form a \emph{Gaussian state}, and sometimes we say that the protocol semantics produces a Gaussian state (as part of a CRN state).
\end{mydef}


The semantics of Definition \ref{GaussSem} combines the end states of sub-protocols by linear operators, as we show in the examples below. We stress that it does not, however, provide a time-domain GP: just the protocol end state as a Gaussian state (random variable). In particular, a protocol like $let\, x\, =\, Equilibrate(P,p)\, in\, Dispose(x)$ introduces a discontinuity at the end time $p$, where all the concentrations go to zero; other discontinuities can be introduced by $Mix$. A protocol may have a finite number of such discontinuities corresponding to liquid handling operations, and otherwise proceeds by LNA-derived GPs and by linear combinations of Gaussian states at the discontinuity points.

It is instructive to interpret our protocol operators as linear operators on Gaussian states: this justifies how covariance matrices are handled in Definition \ref{GaussSem}, and it easily leads to a generalized class of possible protocol operators. We discuss linear protocol operators in the examples below.

\begin{exampleC}
The $Mix$ operator combines the concentrations, covariances, and temperatures of the two input samples proportionally to their volumes. The resulting covariance matrix, in particular, can be justified as follows.
Consider two (input) states $A = (\mu_A,\Sigma_A,V_A,T_A), B = (\mu_B,\Sigma_B,V_B,T_B)$ with $|\mu_A| = |\mu_B| = k$ and $|\Sigma_A| = |\Sigma_B| = k\times k$. Let $0$ and $1$ denote null and identity vectors and matrices of size $k$ and $k\times k$. Consider a third null state $C = (0,0,V_C,T_C)$, with $V_C = V_A + V_B$ and $T_C = \frac{V_AT_A+V_BT_B}{V_C}$. The joint Gaussian distribution of $A,B,C$ is given by
$\mu = 
\begin{bmatrix}
\mu_A \\
\mu_B \\
0
\end{bmatrix}, $
$\Sigma = 
\begin{bmatrix}
\Sigma_A & 0 & 0 \\
0 & \Sigma_B & 0 \\
0 & 0 & 0
\end{bmatrix}. $
Define the symmetric hollow linear operator $Mix = 
\begin{bmatrix}
0 & 0 & \frac{V_A}{V_C}1 \\
0 & 0 & \frac{V_B}{V_C}1 \\
\frac{V_A}{V_C}1 & \frac{V_B}{V_C}1 & 0
\end{bmatrix}.$
The zeros on the diagonal (hollowness) imply that the inputs states are zeroed after the operation, and hence discarded.
Applying this operator to the joint distribution, by the linear combination of normal random variables we obtain a new Gaussian distribution with:
\begin{align*}
&\mu' = Mix \cdot \mu =
\begin{bmatrix}
0 \\
0 \\
\frac{V_A\mu_A + V_B\mu_B}{V_C}
\end{bmatrix} \\
&\Sigma' = Mix \cdot \Sigma \cdot Mix^{\top} = 
\begin{bmatrix}
0 & 0 & 0 \\
0 & 0 & 0 \\
0 & 0 & \frac{V_A^2\Sigma_A + V_B^2\Sigma_B}{V_C^2}
\end{bmatrix}
\end{align*}
Hence, all is left of $\mu',\Sigma'$ is the output state $C = (\mu_C, \Sigma_C, V_C, T_C)$ where $\mu_C = \frac{V_A\mu_A + V_B\mu_B}{V_C}$ and $\Sigma_C = \frac{V_A^2\Sigma_A + V_B^2\Sigma_B}{V_C^2}$ as in Definition \ref{GaussSem}.
\end{exampleC}

\begin{exampleC}
The $Split$ operator splits a sample in two parts, preserving the concentration and consequently the covariance of the input sample. The resulting covariance matrix, in particular, can be justified as follows. Consider one (input) state $A = (\mu_A,\Sigma_A,V_A,T_A)$ and two null states $B = (0,0,pV_A,T_A)$, $C = (0,0,(1-p)V_A,T_A)$ with $0 < p < 1$. As above, consider the joint distribution of these three states,
$\mu = 
\begin{bmatrix}
\mu_A \\
0 \\
0
\end{bmatrix},$
$\Sigma = 
\begin{bmatrix}
\Sigma_A & 0 & 0 \\
0 & 0 & 0 \\
0 & 0 & 0
\end{bmatrix}$
and define the symmetric hollow linear operator $Split = 
\begin{bmatrix}
0 & 1 & 1 \\
1 & 0 & 0 \\
1 & 0 & 0
\end{bmatrix}.$
The 1's imply that concentrations are not affected. The whole submatrix of output states being zero implies that any initial values of output states are ignored. Applying this operator to the joint distribution we obtain:
\begin{align*}
&\mu' = Split \cdot \mu =
\begin{bmatrix}
0 \\
\mu_A \\
\mu_A
\end{bmatrix} \\
&\Sigma' = Split \cdot \Sigma \cdot Split^{\top} = 
\begin{bmatrix}
0 & 0 & 0 \\
0 & \Sigma_A & \Sigma_A \\
0 & \Sigma_A & \Sigma_A
\end{bmatrix}
\end{align*}
By projecting $\mu',\Sigma'$ on $B$ and $C$ we are left with the two output states $B = (\mu_A, \Sigma_A, pV_A, T_A)$, $C = (\mu_A, \Sigma_A, (1-p)V_A, T_A)$, as in Definition \ref{GaussSem}. The correlation between $B$ and $C$ that is present in $\Sigma'$ is not reflected in these outcomes: it is discarded on the basis of the well-mixed-solution assumption.
\end{exampleC}

\begin{exampleC}
In general, any symmetric hollow linear operator, where the submatrix of the intended output states is also zero, describes a possible protocol operation over samples. As a further example, consider a different splitting operator, $let\, x,y = Osmo(P_1,p)\, in\, P_2$, that intuitively works as follows. A membrane permeable only to water is placed in the middle of an input sample $A = (\mu_A, \Sigma_A, V_A, T_A)$, initially producing two samples of volume $V_A/2$ but with unchanged concentrations. Then an osmotic pressure is introduced (by some mechanism) that causes a proportion $0 < p < 1$ of the water (and volume) to move from one side to the other, but without transferring any other molecules. As the volumes change, the concentrations increase in one part and decrease in the other. Consider, as in $Split$, two initial null states $B$ and $C$ and the joint distributions of those three states $\mu,\Sigma$. Define a symmetric linear operator $Osmo = 
\begin{bmatrix}
0 & 1-p & p \\
1-p & 0 & 0 \\
p & 0 & 0
\end{bmatrix}. $
Applying this operator to the joint distribution we obtain:
\begin{align*}
&\mu' = Osmo \cdot \mu =
\begin{bmatrix}
0 \\
(1-p)\mu_A \\
p\mu_A
\end{bmatrix} \\
&\Sigma' = Osmo \cdot \Sigma \cdot Osmo^{\top} = 
\begin{bmatrix}
0 & 0 & 0 \\
0 & (1-p)^2\Sigma_A & p(1-p)\Sigma_A \\
0 & p(1-p)\Sigma_A & p^2\Sigma_A
\end{bmatrix}
\end{align*}
producing the two output states $B = ((1-p)\mu_A, (1-p)^2\Sigma_A, pV_A, T_A)$ and $C = (p\mu_A,$ $p^2\Sigma_A, (1-p)V_A, T_A)$ describing the situation after the osmotic pressure is applied. Again we discard the correlation between $B$ and $C$ that is present in $\Sigma'$ on the basis of the well-mixed-solution assumption.
\end{exampleC}

\begin{exampleC}
We compute the validity of some simple equivalences between protocols by applying the Gaussian Semantics to both sides of the equivalence. Consider the CRN state $Poisson(k,V,T) = ([k], [[k]], V, T)$ over a single species having mean $[k]$ (a singleton vector) and variance $[[k]]$ (a $1\times 1$ matrix). This CRN state can be produced by a specific CRN \cite{laurenti2018molecular}, but here we add it as a new primitive protocol and extend our semantics to include $[\![ Poisson(k,V,T)]\!]^{\rho}=([k], [[k]], V, T)$. Note first that mixing two uncorrelated Poisson states does not yield a Poisson state (the variance changes to $[[k/2]]$). Then the following equation holds:
\begin{align*}
&[\![ Mix(Poisson(k,V,T),Poisson(k,V,T))]\!]^{\rho} \\ = \quad 
&[\![ let \, x,y = Split(Poisson(k,V,T),0.5)\, in\,  Mix(x,y)]\!]^{\rho} \\ = \quad 
&([k], [[k/2]], V, T)
\end{align*}
That is, in the intrinsic-noise semantics, mixing two correlated (by Split) Poisson states is the same as mixing two uncorrelated Poisson states, because of the implicit decorrelation that happens at each $Split$. An alternative semantics that would take into account a correlation due to $Split$ could instead possibly satisfy the appealing equation $[\![ let \, x,y = Split(P,0.5)\, in\,  Mix(x,y)]\!]^{\rho}=[\![P]\!]^{\rho}$ for any $P$ (i.e., splitting and remixing makes no change), but this does not hold in our semantics for $P = Poisson(k,V,T)$, where the left hand side yields $([k], [[k/2]], V, T)$ (i.e., splitting, decorrelating, and remixing makes a change).

The $Mix$ operator can be understood as diluting the two input samples into the resulting larger output volume. We can similarly consider a $Dilute$ operator that operates on a single sample, and we can even generalize it to concentrate a sample into a smaller volume. In either case, let $W$ be a new forced volume, and $U$ be a new forced temperature for the sample. Let's define $[\![ Dilute(P,W,U)]\!]^{\rho}= (\frac{V}{W}\mu,\frac{V^2}{W^2}\Sigma,W,U)$
where $(\mu,\Sigma,V,T)= [\![P]\!]^{\rho}$. Then the following equation holds: 
\begin{align*}
&[\![ Mix(Dilute(P_1,W_1,U_1), Dilute(P_2,W_2,U_2))]\!]^{\rho} \\ = \quad
&[\![ Dilute(Mix(P_1,P_2),W_1+W_2,\frac{W_1U_1 + W_2U_2}{W_1+W_2})]\!]^{\rho} \\ = \quad
&(\frac{V_1\mu_1 + V_2\mu_2}{W_1+W_2}, \frac{V_1^2\Sigma_1 + V_2^2\Sigma_2}{(W_1+W_2)^2}, W_1+W_2, \frac{W_1U_1 + W_2U_2}{W_1+W_2})
\end{align*}
Note that, by diluting to a fixed new volume W, we obtain a protocol equivalence that does not mention, in the equivalence itself, the volumes $V_i$ resulting from the sub-protocols $P_i$ (which of course occur in the semantics). If instead we diluted by a factor (e.g. factor=2 to multiply volume by 2), then it would seem necessary to refer to the $V_i$ in the equivalence.
\end{exampleC}

To end this section, we provide an  extension of syntax and semantics for the optimization of protocols. An \emph{optimize} protocol directive identifies a vector of variables within a protocol to be varied in order to minimize an arbitrary cost function that is a function of those variables and of a collection of initial values. The semantics can be given concisely but implicitly in terms of an $argmin$ function: a specific realization of this construct is then the subject of the next section.

\begin{mydef}{(Gaussian Semantics of a Protocol - Optimization)}\label{def:OptimizationSem}

Let $z \in Par^N$ be a vector of (optimization) variables, and $k \in Par^M$ be a vector of (initialization) variables with (initial values) $r \in \mathbb{R}^{M}$, where all the variables are distinct and $Par$ is disjoint from $Var$. Let $P \in Prot$ be a protocol for a CRN $\mathcal{C}=(\mathcal{A},\mathcal{R})$, with free variables $z$ and $k$ and with any other free variables covered by an environment $\rho \in Env = (Var \cup Par) \to S$. Given a cost function $C : \mathbb{R}^{|\mathcal{A}|}\times \mathbb{R}^{N}\to \mathbb{R}$, and a dataset $\mathcal{D} \subseteq_{fin} \mathbb{R}^M \times \mathbb{R}^N \times \mathbb{R}^{|\mathcal{A}|}$, the instruction ``optimize k=r, z in P with C given $\mathcal{D}$" provides a vector in $\mathbb{R}^N$ of optimized values for the $z$ variables. (For simplicity, we omit extending the syntax with new representations for $C$ and $\mathcal{D}$, and we let them represent themselves). The optimization is based on a stochastic process $\mathbf{X} \in \mathbb{R}^{M+N} \rightarrow \mathbb{R}_{\geq 0}^{|\mathcal{A}|}\times\mathbb{R}^{|\mathcal{A}|\times|\mathcal{A}|}$ derived from $P$ via Definition \ref{GaussSem}.
\begin{align*}
     &[\![optimize\, k=r,\, z\, in\, P\, with\, C\, given\, \mathcal{D}]\!]^{\rho}=\\
      &\quad \quad argmin_{u \in \mathbb{R}^N} \, \mathbb{E}_{y \sim \mathbf{X}(r,u|\mathcal{D})}[ C(y,u) ] \\
      &\quad \quad where \, \mathbf{X}(r,u) = [\![P]\!]^{\rho\{k \leftarrow r, z\leftarrow u\}} \, for\,all \, r \in \mathbb{R}^{M} \, and \, u \in \mathbb{R}^{N},
\end{align*}
where $\mathbb{E}_{y \sim \mathbf{X}(r,u|\mathcal{D})}[ \cdot ]$ stands for the expectation with respect the conditional distribution of $\mathbf{X}(r,u)$ given $\mathcal{D}$.
\end{mydef}

From Definition \ref{GaussSem} it follows that, 
for any $(r,u)\in \mathbb{R}^{M+N},$
$\mathbf{X}(r,u)$ is a Guassian random variable. In what follows, for the purpose of optimization, we further assume that $\mathbf{X}$ is a Gaussian process defined on the sample space $\mathbb{R}^{M+N}$, i.e., that for any possible $(r_1,u_1)$ and $(r_2,u_2),$ $\mathbf{X}(r_1,u_1)$ and $\mathbf{X}(r_2,u_2)$ are jointly Gaussian. Such an assumption is standard \cite{rasmussen2003gaussian} and natural for our setting (e.g., it is guaranteed to hold for different equilibration times) and guarantees that we can optimize protocols by employing results from Gaussian process optimization as described in detail in Section \ref{sec:Optimization}. 

\subsection{Optimization of Protocols through Gaussian Process Regression}
\label{sec:Optimization}
\noindent
In Definition \ref{def:OptimizationSem} we introduced the syntax and semantics for the optimization of a protocol from data. In this section, we show how the optimization variables can be selected in practice. In particular, we leverage existing results for Gaussian processes (GPs) \cite{rasmussen2003gaussian}. 
We start by considering the dataset $\mathcal{D}=\{(r_i,u_i,y_i), i\in \{1,...,n_{\mathcal{D}}\} \} \subseteq_{fin} \mathbb{R}^M \times \mathbb{R}^N \times \mathbb{R}^{|\mathcal{A}|}$ comprising  $n_{\mathcal{D}}$ executions of the protocol for possibly different initial conditions, i.e., each entry gives the concentration of the species after the execution of the protocol $(y_i)$, where the protocol has been run with $k=r_i, z=u_i$. 
Then, we can predict the output of the protocol starting from $\bar{x}=(r,u)$ by computing the conditional distribution of $\mathbf{X}$ (Gaussian process introduced in Definition \ref{def:OptimizationSem}) given the data in  $\mathcal{D}$.  In particular, under the assumption that $\mathbf{X}$ is Gaussian, it is well known that the resulting posterior model,  $\mathbf{X}(\bar{x}|\mathcal{D})$, is still Gaussian  with mean and covariance functions given by 
\begin{align}
   \label{Eq:COnditionMean}& \bm{\mu}_p(\bar{x})= \bm{\mu}(\bar{x}) + \bm{\Sigma}_{\bar{x},\mathcal{D}}(\bm{\Sigma}_{\mathcal{D},\mathcal{D}}+\sigma^2 I)^{-1}({y}_{\mathcal{D}}-\bm{\mu}_{\mathcal{D}})\\
   &{\bm{\Sigma}}_p(\bar{x},\bar{x})= \bm{\Sigma}_{\bar{x},\bar{x}}-\bm{\Sigma}_{\bar{x},\mathcal{D}}(\Sigma_{\mathcal{D},\mathcal{D}}+\sigma^2 I)^{-1}\bm{\Sigma}_{\bar{x},\mathcal{D}}^T,\label{Eq:ConditionalVariance}
\end{align} 
where $\sigma^2 I$ is a diagonal covariance modelling i.i.d. Gaussian observation noise with variance $\sigma^2$, $\bm{\mu}(\bar{x})$ and $\bm{\Sigma}_{\bar{x},\bar{x}}$ are the prior mean and covariance functions, $\bm{\Sigma}_{\mathcal{D},\bar{x}}$ is the covariance between $\bar{x}$ and all the points in $\mathcal{D}$, and $y_{\mathcal{D}}$, $\bm{\mu}_{\mathcal{D}}$ are vectors of dimensions $\mathbb{R}^{|\Lambda|\cdot n_{\mathcal{D}}}$ containing for all $(x_i,u_i,y_i)\in \mathcal{D}$ respectively $y_i$ and  $\bm{\mu}((x_i,u_i))$ \cite{rasmussen2003gaussian}. 
Note that, in order to encode our prior information of the protocol, we take $\bm{\mu}(\bar{x})$ to be the mean of the GP as defined in Definition \ref{GaussSem}, while for the variance we can have $\bm{\Sigma}_{\bar{x},\bar{x}}$ to be any standard kernel \cite{rasmussen2003gaussian}; in the experiments, as is standard in the literature, we consider the widely used squared exponential kernel, which is expressive enough to approximate any continuous function arbitrarily well \cite{micchelli2006universal}. However, we remark that,  for any parametric kernel, such as the squared exponential, we can still select the hyper-parameters that best fit the variance given by Definition \ref{GaussSem} as well as the data \cite{rasmussen2003gaussian}. We should also stress that the resulting $\bm{X}(\bar{x}|\mathcal{D})$ is a Gaussian process that merges in a principled way (i.e., via Bayes' rule) our prior knowledge of the model (given by Definition \ref{GaussSem}) with the new information given in $\mathcal{D}.$ Furthermore, it is worth stressing again that $\mathbf{X}(\bar{x}|\mathcal{D})$ is a GP with input space given by $\mathbb{R}^{M+N},$ which is substantially different from the GP defined by the LNA (Definition \ref{LangSynt2}), where the GP was defined over time.


For a given set of initialization parameters $r$ our goal is  to synthesize the optimization variables that optimize the protocol with respect to a given cost specification $C : \mathbb{R}^{|\mathcal{A}|}\times \mathbb{R}^{N}\to \mathbb{R}$,  i.e., we want to find $u^*$ such that 
\begin{align}
    \label{cost}
u^* = argmin_{u \in \mathbb{R}^N} \mathbb{E}_{y  \sim \mathcal{N}(\bm{\mu}_p(\bar{x}),{\bm{\Sigma}}_p(\bar{x},\bar{x}))}[ C(y,u) ],  
\end{align}
\noindent
where $\mathbb{E}_{y \sim \mathcal{N}(\bm{\mu}_p(\bar{x}),{\bm{\Sigma}}_p(\bar{x},\bar{x}))}[ \cdot ]$ is the expectation with respect to the GP given by Eqn \eqref{Eq:COnditionMean} and \eqref{Eq:ConditionalVariance}). Note that $r$ is a known vector of reals (see Definition \ref{def:OptimizationSem}), hence we only need to optimize for the free parameters $u$. 
In general, the computation of $u^*$ in Eqn \eqref{cost} requires solving a non-convex optimization problem that cannot be solved exactly \cite{boyd2004convex}.  In this paper we approximate Eqn \eqref{cost} via gradient-based methods. These methods, such as gradient descent, require the computation of the gradient of the expectation of $C$ with respect to $u$:
\begin{align}
    \label{Eqn:DerivativeCost}
    \frac{ \partial \mathbb{E}_{y \sim \mathcal{N}(\bm{\mu}_p(\bar{x}),{\bm{\Sigma}}_p(\bar{x},\bar{x}))}[ C(y,u) ]}{\partial u}.
\end{align} 
Unfortunately, direct computation of the gradient in \eqref{Eqn:DerivativeCost} is infeasible in general, as the probability distribution where the expectation is taken depends on $u$ itself (note that $\bar{x}=(r,u)$). However, for the GP case, as shown in Lemma \ref{Lemma:GradientCostExp}, the gradient of interest can be computed directly by  reparametrizing the  Gaussian distribution induced by Eqn \eqref{Eq:COnditionMean} and \eqref{Eq:ConditionalVariance}.
\begin{lemmaC}
\label{Lemma:GradientCostExp}
For $\bar x=(r,u)$, let  $D(\bar{x})$ be the matrix such that $D(\bar{x}) D^T(\bar{x})={\bm{\Sigma}}_p(\bar{x},\bar{x})$. Then, it holds that
\begin{equation}
    \label{Eqn:Derivative}
    \frac{ \partial \mathbb{E}_{y \sim \mathcal{N}(\bm{\mu}_p(\bar{x}),{\bm{\Sigma}}_p(\bar{x},\bar{x}))}[ C(y,u) ]}{\partial u}=    \mathbb{E}_{z \sim \mathcal{N}(0,I)}[\frac{ \partial C(\bm{\mu}_p(\bar{x})+D(\bar{x}) z,u)}{\partial u} ].
\end{equation}
\end{lemmaC}
\noindent
In particular, $D(\bar{x})$ as defined above is guaranteed to exist under the assumption that ${\bm{\Sigma}}_p$ is positive definite and can be computed via Cholesky decomposition. Furthermore, note that if the output is uni-dimensional then $D(\bar{x})$ is simply the standard deviation of the posterior GP in $\bar{x}$.


\begin{exampleC}
Given the CRN $\mathcal{C}$ introduced in Example \ref{RunningExample} consider the protocol
$$ P=Equilibrate((0.1 mM ,0.001 mM,0.001 mM,1 \mu L,20 C),T), $$
which seeks to evolve the various species from an initial starting concentration of $0.1$ mM for species $a$ and $0.001$ mM for both species $b$ and $c$ for $T$ seconds.
Hence, for this example we have $r=[0.1,0.001,0.001,1,20]$ including initial conditions, volume, and temperature and $u=[T]$, that is, the only optimization variable is the equilibration time.
We consider a dataset $\mathcal{D}$ composed of the following six data points (for simplicity we report the output values of only species $b$ and omit unit of measurement, which are mM for concentration, $\mu$L for volume, Celsius degrees for temperature, and seconds for equilibration time):
\begin{align*}
   & d_1=((0.1001, 0.0015, 0.001,1,20),0,0.001)\quad d_2=((0.099, 0.001, 0.001,1,20),40,0.001)\\
   &d_3=((0.1, 0.001, 0.001,1,20),150,0.09) \quad d_4=((0.1, 0.001, 0.002,1,20),250,0.08) \\ 
   &d_5=((0.09, 0.001, 0.0015,1,20),400,0.003) \quad d_6=((0.1, 0.001, 0.002,1,20),500,0.001),
\end{align*}
where, for example, in $d_1$ we have that 
the initial concentration for $a,b,c$ are respectively $0.1001, 0.0015, 0.001$, volume and temperature at which the experiment is performed are $1$ and $20$ and the observed value for $b$ at the end of the protocol for $T=0$ is $0.001.$ We assume an additive Gaussian observation noise (noise in the collection of the data) with standard deviation $\sigma=0.01$.
Note that the initial conditions of the protocol in the various data points in general differ from those of $P$, which motivates having Eqn \eqref{Eq:COnditionMean} and \eqref{Eq:ConditionalVariance} dependent on both $r$ and $u$.

We consider the prior mean given by Eqn \ref{Eqn:Mean} and independent squared exponential kernels for each species with hyper-parameters optimized through maximum likelihood estimation (MLE) \cite{rasmussen2003gaussian}. The resulting posterior GP is reported in Figure \ref{fig:ExRunnwithGPcost}, where is compared with the prior mean and the true underlying dynamics for species $b$ (assumed for this example to be deterministic). We note that with just $6$ data points in $\mathcal{D}$ the posterior GP is able to correctly recover the true behavior of $b$ with relatively low uncertainty.

\begin{figure}
\centering
\includegraphics[width=1\linewidth]{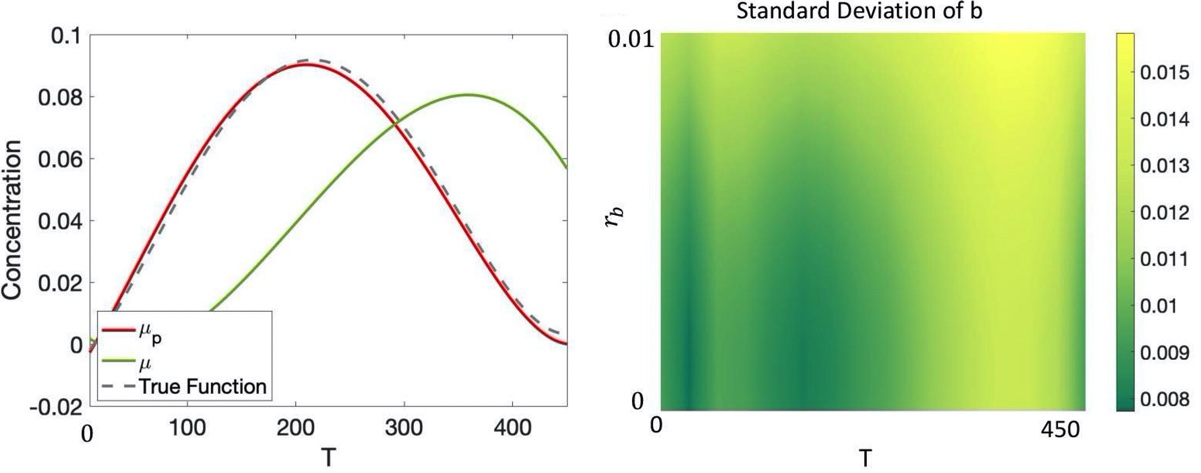}
\caption{
Mean and variance of species $b$ for the CRN reported in Example \ref{RunningExample} for $r=[0.1,r_b,0.001,1,20]$ and $u=[T].$ \textbf{Left:} Evolution of $\mu$ (prior mean of species $b$ given by Eqn \eqref{Eqn:Mean}), $\bm{\mu}_p$ (posterior mean of species $b$ given by Eqn \eqref{Eq:COnditionMean}), and the true dynamics of species $b$, assumed to be a deterministic function for this example, for $r_b=0.001$ (initialization variable relative to species $b$). It is possible to observe how, with just a few data points, the posterior mean {reflects} correctly the true dynamics. \textbf{Right:} Standard deviation of $b$ after training (square root of solution of Eqn \eqref{Eq:ConditionalVariance}) as a function  of $r_b$  and $T$. The variance is higher for combinations of $T$ and $r_b$ where no training data are available. 
} 
\label{fig:ExRunnwithGPcost}
\end{figure}

We would like to maximize the concentration of $b$ after the execution of the protocol. This can be easily encoded with the following cost function:
$$ C(a,b,c,u)= -b^2.$$
 For $r=[0.1,0.001,0.001,1,20]$ the value obtained is $T\sim 230$, which is simply the one that maximizes the posterior mean. In Section \ref{sec:ExpGIbson} we will show how, for more complex specifications, the optimization process will become more challenging because of the need to also account for the variance in order to balance between exploration and exploitation.

\end{exampleC}

\section{Results}

\noindent
We consider two case studies where we illustrate the usefulness of our framework. The first example illustrates how our framework can be employed to optimize a Gibson assembly protocol \cite{gibson2009enzymatic} from experimental data.
 In the second example we illustrate the flexibility of our language and semantics on a combination of protocol operations. Furthermore, we also use this example to show how our framework can be employed to perform analysis of the protocol parameters while also accounting for the uncertainty in both the model dynamics, the protocol parameters, and the  data collection. 

\subsection{Gibson Assembly Protocol}
\label{sec:ExpGIbson}
\noindent
We start by considering a simplified version of the Gibson assembly, a protocol widely used for  joining  multiple DNA fragments \cite{gibson2009enzymatic}. 
We consider the following model of Gibson assembly described by Michaelis-Menten kinetics, where two DNA double strands, AB and BA (with shared endings A and B but different middle parts), are enzymatically joined in two symmetric ways according to their common endings, and the resulting strands ABA and BAB are circularized into the same final structure O. The resulting dynamical model is given by the following ODEs (assuming AB is over abundant with respect to BA): 
\begin{align}
    & \label{Eqn:FirstKineticsGIbson}\frac{d AB(t)}{dt}=  0\\
    & \frac{d BA(t)}{dt}=-\frac{k_{cat_1}\cdot AB(t)\cdot BA(t)}{BA(t) +K_{m_1} } - \frac{k_{cat_2}\cdot AB(t)\cdot BA(t)}{BA(t) +K_{m_2} } \\
    & \frac{d ABA(t)}{dt}= \frac{k_{cat_1}\cdot AB(t)\cdot BA(t)}{BA(t) +K_{m_1} } -\frac{k_{cat_1}\cdot ABA(t)}{ABA(t) +K_{m_1} }  \\
    & \frac{d BAB(t)}{dt}= \frac{k_{cat_2}\cdot AB(t)\cdot BA(t)}{BA(t) +K_{m_2} } -\frac{k_{cat_2}\cdot BAB(t)}{BAB(t) +K_{m_2} } \\
 & \label{Eqn:LastKineticsGIbson}\frac{d O(t)}{dt}=\frac{k_{cat_1}\cdot ABA(t)}{ABA(t) +K_{m_1} } + \frac{k_{cat_2}\cdot BAB(t)}{BAB(t) +K_{m_2} } ,
\end{align}
where $k_{cat_1}, k_{cat_2}, K_{m_1},K_{m_2}  $ are given parameters. Gibson assembly is a single-step isothermal protocol: once all the ingredients are combined, it can be written as:
$$ P=Equilibrate(([1 mM,x_{BA} mM,0mM,0mM,0mM],1\mu L,50 C),T),  $$
where $[1,x_{BA},0,0,0]$ is a vector of the initial concentration of the various species with $AB$ initialized at 1 mM, $BA$ at $x_{BA}$ mM, and all the other species to $0$ mM for a sample of volume of $1$ $\mu L$  and at a temperature $50$ Celsius degrees, equilibrated for $T$ seconds.  For this example we have that the set of optimization variables is $u=[x_{BA},T]$ and the goal is to find initial condition of $BA$ ($x_{BA}$) and equilibration time ($T$) such that,  at the end of the protocol, species $O$ has a concentration as close as possible to a desired value $O_{des},$ while also keeping the equilibration time and the initial concentration of $BA$ close to some reference values. This can be formalized with the following cost function: 
$$ C(x_{BA},T )=\big(\beta (O(T)-O_{des})\big)^2 + \lambda (\frac{T-T_{ref}}{T_{norm}})^2 + (1-\lambda) (x_{BA}-I_{ref})^2,  $$
where $O(T)$ is the concentration of strand $O$ after the execution of the protocol. $T_{ref},$ $T_{norm},$ $I_{ref},$ and $\beta,$  are parameters describing, respectively, the reference equilibration time, a normalization term for the equilibration time,  the reference initial concentration of $BA$, and a weight term such that if $\beta>1$ we give more importance to $O(T)$ being close to $O_{des}$ compared to equilibration time and initial conditions being closer to their reference values. Similarly, $\lambda\in [0,1]$ is a weight term that balances the importance that  equilibration time and the initial concentration of $BA$ are close to their reference values.   In our experiments we fix $O_{des}=0.6,$ $T_{ref}=700,$ $T_{norm}=1000,$ $I_{ref}=1$, and $\beta=20.$

The dynamical model considered in Eqn \eqref{Eqn:FirstKineticsGIbson}-\eqref{Eqn:LastKineticsGIbson} is obtained by combining multiple enzymatic reactions into single Michaelis-Menten steps and the value of the reaction rates may not be accurate. Hence, to obtain a more accurate model, 
we combine Eqn \eqref{Eqn:FirstKineticsGIbson}-\eqref{Eqn:LastKineticsGIbson} with the experimental data from \cite{gibson2009enzymatic}  under the assumption that observations are corrupted by a Gaussian noise of  relatively high standard deviation of $0.1.$ 
\begin{figure}
\centering
\includegraphics[width=1\linewidth]{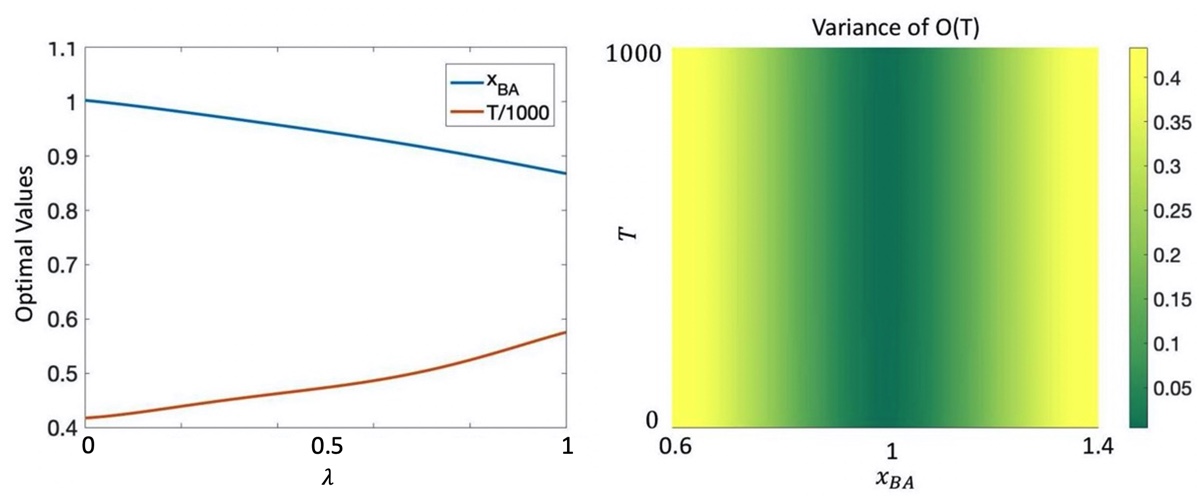}
\caption{
Optimal values for $x_{BA}$ and ${T}$ and variance of the predictive Gaussian process. \textbf{Left:}  Optimal values of $x_{BA}$ and $\frac{T}{1000}$ for different values of $\lambda.$ \textbf{Right:}  Variance of $O(T)$ after training (Eqn \eqref{Eqn:Variance}) as a function  of $x_{BA}$ and $T$. The variance is minimized for $x_{BA}\sim 1$. This is due to the fact that all training data  have $x_{BA}=1$. 
}
\label{fig:GibosnOptimalValues}
\end{figure}
In Figure \ref{fig:GibosnOptimalValues} we plot the  synthesized (approximately) optimal values of $T$ and $x_{BA}$ for various values of $\lambda$ by using gradient descent, with the gradient computed as shown in Lemma \ref{Lemma:GradientCostExp}.  Note that even for $\lambda=0$ or $\lambda=1$ the cost cannot be made identically zero. This is due to the uncertainty in the model that leads to a non-zero variance everywhere. Note also that for $\lambda=1$ the synthesized time $T$ is smaller than $700$ (the reference equilibration time). This can be explained by looking at Figure \ref{fig:GibosnOptimalValues} (Right), where we plot the variance of $O(T)$ as a function of $T$ and $x_{BA}$. In fact, as we have available only data (reported in the Appendix B) for $x_{BA}\sim 1$ the variance increases when $x_{BA}$ is far from $1$. As a consequence, the algorithm automatically balances this tradeoff by picking an equilibration time that is close to the target value but also allows for an $x_{BA}$ close to $1$ in order the keep the variance under control.

\subsection{Split and Mix Protocol}
\label{sec:Split and Mix}
\noindent
Consider the CRN $\mathcal{C}=(\{a,b,c\},\mathcal{R})$, where $\mathcal{R}$ is given by the reactions:
$$a + b \rightarrow^{1} b + c \quad \quad \quad 
a + c \rightarrow^{1} a + a \quad \quad \quad 
b + c \rightarrow^{1} c + c. $$
Consider also the associated protocol $P_{split\&mix}$ shown below in the syntax of Definition \ref{LangSynt2} (with underscore standing for an unused variable, with initial concentration vectors ordered as $(a_0, b_0, c_0)$, and initial states matching the pattern $((a_0, b_0, c_0)$, $volume$, $temperature)$: 
\begin{align*}
& let\, A = ((10mM, 0mM, 1mM), 1\mu L, 20C)\, in\\
& let\,  A1 = Equilibrate(A,100s)\, in\\
& let\,  C,D = Split(A1,0.5)\, in\\
& let\,\,  \_ = Dispose(C)\, in\\
& let\,  B = ((0mM, 10mM, 1mM), 1\mu L, 20C)\, in\\
& let\,  B1 = Equilibrate(B,100s)\, in\\
& let\,  E = Mix(D,B1)\, in\\
& Equilibrate(E,1000s)
\end{align*}

\begin{figure}
\centering
\includegraphics[width=1\linewidth]{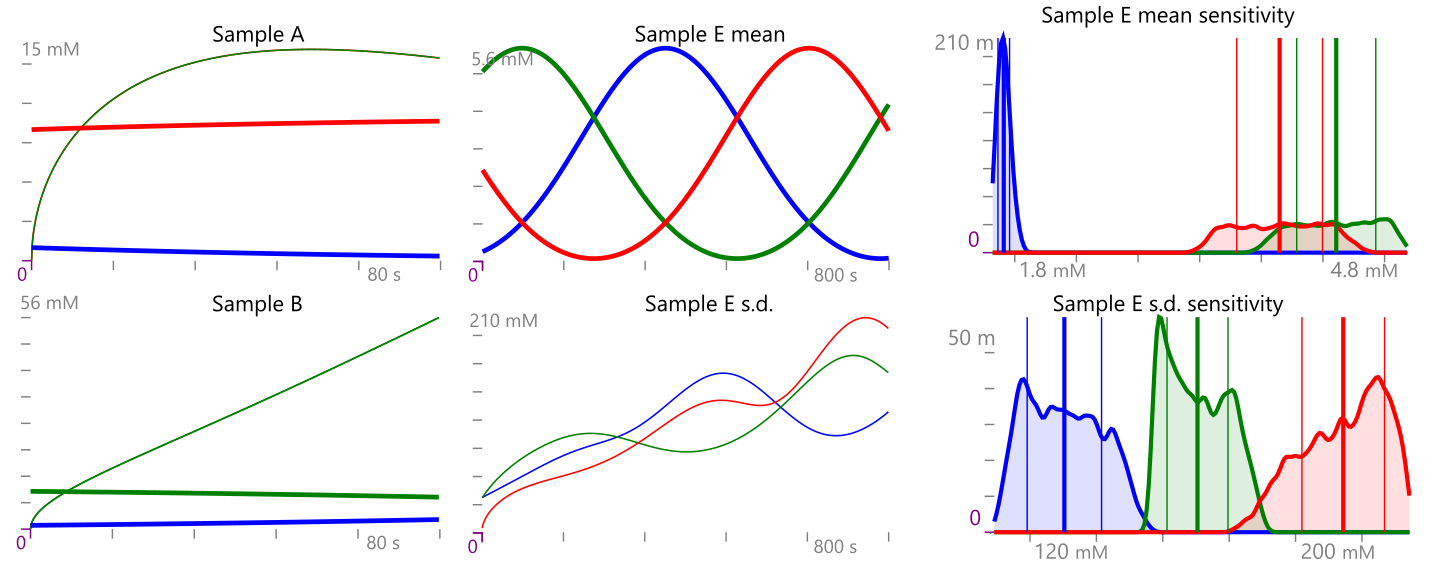}
\caption{
\textbf{Left \& Center:} Evolution of $a$ (red), $b$ (green), $c$ (blue) for protocol $P_{split\&mix}$, showing mean (thick lines) and standard deviation (thin lines), separately for Sample E, with some trajectory overlaps in Sample A and B. Horizontal axis is time ($s$), vertical axis is concentration ($mM$). Sample A is simulated first, then Sample B, and finally Sample E, where the standard deviations start above zero due to propagating the final states of the earlier simulations. \textbf{Right:} Density plots for global sensitivity analysis over 3000 runs, displaying the sensitivity at the end of the protocol of mean (top) and standard deviation (bottom) of $a$, $b$, $c$. Sensitivity is with respect to the three $Equilibrate$ duration parameters and the $Split$ proportion parameter: those prameters are simultaneously drawn from uniform distributions, each varying by up to $\pm$5\%. Horizontal axis is concentration ($mM$), vertical axis is the kernel density estimate ($m = \times 10^{-3}$) with a standard normal distribution kernel, and bandwidth of 1/100 of the data range. Thick vertical lines locate the mean, thin vertical lines locate the $\pm$ standard deviation. 
}
\label{fig:SplitAndMix}
\end{figure}


This protocol interleaves equilibration steps with mixing and splitting. During the first $Equilibrate$, only the second reaction is active in sample $A$, due to its initial conditions, yielding monotonic changes in concentrations. During the second $Equilibrate$, similarly, only the third reaction is active in sample $B$. During the third $Equilibrate$, after samples $A$ and $B$ have been manipulated and mixed, all three reactions are active in sample $E$, yielding an oscillation. 

The semantics of Definition \ref{GaussSem} can be used to unravel the behavior of $P_{split\&mix}$. In particular, integration of the CRN $\mathcal{C}$ in samples A and B yields states $S_A$, $S_B$ that, after some split and mix manipulations, determine the initial conditions for both mean and covariance for the integration of $\mathcal{C}$ in sample E.

We can numerically evaluate the protocol following the semantics: this is shown in 
Figure \ref{fig:SplitAndMix} (Left \& Center), where the protocol evaluation results in three simulations that implement the three integration steps, where each simulation is fed the results of the previous simulations and protocol operations. 


To show the usefulness of the integrated semantics, in Figure \ref{fig:SplitAndMix} (Right) we perform a global sensitivity analysis of the whole protocol with respect to the protocol parameters $t_1=100$, $t_2=100$, $t_3=1000$, the duration of the three $Equilibrate$, and $s_1=0.5$, the proportion of $Split$. We produce density plots of the means and variances of $a$,$b$,$c$ at the end of the protocol when $t_1$,$t_2$,$t_3$,$s_1$ vary jointly randomly by up to 10\%. This is thus a sensitivity analysis of the joint effects of three simulations connected by liquid handling steps. It is obtained by providing the whole parameterized protocol, not its separate parts, to the harness that varies the parameters and plots the results. We could similarly vary the initial concentrations and reaction rates, and those together with the protocol parameters. 


\section{Discussion}

{ 

Automation already helps scaling up the production of chemical and biochemical substances, but it is also hoped it will solve general reproduceability problems. In the extreme, even one-off experiments that have no expectation of reproduction or scaling-up should be automated, so that the provenance of the data they produce can be properly recorded. Most pieces of lab equipment are already computer controlled, but automating their interconnection and integration is still a challenge that results in poor record keeping. A large amount of bottom-up development will be necessary to combine all these pieces of equipment into "fabrication lines". But it is also important to start thinking top-down at what the resulting connecting "glue" should look like, as we have attempted to do in this paper. This is because, if automation is the ultimate goal, then some pieces of equipment that are hard to automate should be discarded entirely.

Consider, for example, the Split operation from Definition \ref{GaussSem}, which separates a sample into two samples. That represents a vast array of laboratory procedures, from simple pipetting to microfluidic separation, each having its own tolerances and error distributions (which are well characterized and could be included into the language and the analysis). Despite all those variations, it is conceivable that a protocol compiler could take an abstract Split instruction, and a known piece of equipment, and automatically tailor the instruction for that equipment. Digital microfluidics is particularly appealing in this respect because of the relative ease and uniformity of such tailoring \cite{newman2019high}. Therefore, one could decide to work entirely within digital microfluidics, perhaps helping to make that area more robust, and  avoid other kinds of equipment.


Are there drawbacks to this approach? First, as we mentioned, equipment should be selected based on ease of automation: whole new lab procedures may need to be developed in replacement, along with standards for equipment automation. Second, protocol languages are more opaque than either sets of mathematical equations or sequences of laboratory steps. This needs to be compensated for with appropriate user interfaces for common use in laboratories. }

{ 
About the second point, as a first step, we have separately produced an easily deployed app (Kaemika \cite{cardelli2020kaemika}) that supports the integrated language described here, although pragmatic details of the syntax diverge slightly. We have used it to run simulations and to produce Figures \ref{fig:Running} and \ref{fig:SplitAndMix} (see Appendix A). It uses a standard ODE solver for simulating the "Equilibrate" steps of the protocols, i.e., the chemical reaction kinetics. It uses the Linear Noise Approximation (LNA) for stochastic simulations; that is, means and variances of concentrations are computed by an extended set of ODEs and solved with the same ODE solver along with the deterministic rate equations. The liquid handling steps of the protocols are handled following the semantics of Definition \ref{GaussSem}, noting that in the stochastic case means and variances are propagated into and out of successive Equilibrate and liquid handling steps. The result is an interleaved simulation of Equilibrate and liquid handing steps, which is presented as an interleaved graphical rendering of concentration trajectories and of droplet movements on a simulated microfluidic device. The sequence of protocol steps can be extracted as a graph, omitting the detailed kinetics. The kinetic equations operating at each step of the protocol can be extracted as well.

In this context we should also stress that our proposed Gaussian semantics makes various assumptions: (1) we assume that molecular processes described by CRNs and liquid handling operations can be modelled by a Gaussian process, and (2) we assume that the collected data are corrupted by additive Gaussian noise. The former assumption is justified by the precision of lab equipment to perform liquid handling operations, whose uncertainty is generally very small, and by the Central Limit Theorem (CLT). The CLT guarantees that the solution of the Chemical Master Equation will converge to a Gaussian process in the limit of high number of molecules \cite{ethier2009markov}, as is common in wet lab experiments. In particular, as illustrated in Experiment 5.1 in \cite{cardelli2016stochastic}, already a number of molecules of the order of hundreds for each species generally guarantees that a Gaussian approximation is accurate. It is obvious that, in case of experiments with single molecules, such an approximation may be inaccurate  and a full treatment of the CME would be more appropriate \cite{schwabe2012transcription}. 
The assumption that observations are corrupted by Gaussian additive noise is standard \cite{leake2014analytical}. However, we acknowledge that in certain scenarios non-Gaussian or multiplicative observation noise may be required. In this case we would like to stress that Gaussian process regression can still be performed with success, at the price of introducing approximations on the computation of the posterior mean and variance \cite{rasmussen2003gaussian}.

 As automation standards are developed for laboratory equipment, we will be able to target our language to such standards, and extend it or adapt it as needed. In this context, our future works include extension to both the syntax and semantics of our language to include common laboratory procedures that we have not investigated here, for example, changing the temperature of a sample (as in a thermal cycler) or its volume (as in evaporation or dilution). These are easy to add to our abstract framework, but each corresponds to a whole family of lab equipment that may need to be modeled and integrated in detail. Furthermore, we plan to extend the semantics to allow for more general stochastic processes that may be needed when modelling single molecules.

}

{ 
\section{Conclusions}
\noindent
We have introduced a probabilistic framework that rigorously describes {the joint handling of liquid manipulation steps and chemical kinetics}, throughout the execution of an experimental protocol, with particular attention to the propagation of uncertainty and the optimization of the protocol parameters. 
A central contribution of this paper is the distinction between the intrinsic and extrinsic approach to noise, which leads to a much simplified semantics under a chemically justified assumption of well-mixed-solutions. The semantics is reduced to operations on Gaussian states, where any modeled or hypothetical protocol operation is a symmetric linear operator on Gaussian states. 

The Gaussian process semantics approach is novel to this work, and is novel with respect to the Piecewise Deterministic Markov Process semantics in \cite{abate2018experimental}, which treats chemical evolution as deterministic. The semantics in this work is about a collection of deterministic protocol operations, but note that (1) stochasticity in chemical kinetics is propagated across protocol operations, making the whole protocol stochastic, and (2) we have shown examples of how to easily incorporate new protocol operators as linear operators on Gaussian states, which may include ones that introduce their own additional stochasticity. 
The Gaussian approach in this paper enables principled integration of a protocol model with experimental data, which in turn enables automated optimization and analysis of experimental biological protocols.
The syntax of protocols is chosen for mathematical simplicity, reflecting the one in \cite{abate2018experimental}; a closely related but more pragmatic syntax is now implemented in \cite{cardelli2020kaemika}. 

}

\vspace{6pt} 



\authorcontributions{Conceptualization, L.C, M.K., and L.L; methodology, L.C., M.K., and L.L.; software, L.C., and L.L.; validation, L.C., M.K., and L.L.; formal analysis, L.C., M.K., and L.L.; investigation, L.C., M.K., and L.L.; resources, N/A; data curation, N/A; writing---original draft preparation, L.C., M.K., and L.L.; writing---review and editing, L.C., M.K., and L.L.; visualization, L.C. and L.L.; supervision, L.C. and M.K.; project administration, L.C.; funding acquisition, L.C. 
}

\funding{This research was funded in part by the ERC
under the European Union’s Horizon 2020 research and
innovation programme (FUN2MODEL, grant agreement
No. 834115). Luca Cardelli was funded by a Royal Society Research Professorhip RP/R/180001 \& RP/EA/180013.}

\conflictsofinterest{The authors declare no conflict of interest.} 





\appendixtitles{no} 
\appendixstart
\appendix
\section{Simulation Script}

This is the script for the case study of Section \ref{sec:Split and Mix} for the protocol $P_{split\&mix}$, in Kaemika \cite{cardelli2020kaemika}. The protocol begins at `\verb|species {c}|' and ends at `\verb|equilibrate E|'. For the sensitivity analysis of Figure \ref{fig:SplitAndMix}, a function \verb|f| abstracts the \verb|equilibrate| time parameters \verb|e1,e2,e3| and the \verb|split| proportion parameter \verb|s1| and yields the concentrations of \verb|a,b,c| at the end of the protocol.
A multivariate random variable \verb|X|, over a uniform multidimensional sample space \verb|w|, is constructed from \verb|f| to vary the parameters. Then \verb|X| is sampled and plotted.

\begin{verbatim}
function f(number e1 e2 e3 s1) {
define
    species {c}

    sample A 1μL, 20C
    species a @ 10mM in A
    amount c @ 1mM in A
    a + c -> a + a {1}
    equilibrate A1 = A for e1

    sample B {1μL, 20C}
    species b @ 10mM in B
    amount c @ 1mM in B
    b + c -> c + c {1}
    equilibrate B1 = B for e2

    split C,D = A1 by s1
    dispose C

    mix E = D, B1
    a + b -> b + b {1}

    equilibrate E for e3
    
  yield [observe(a,E), observe(b,E), observe(c,E)]
}

random X(omega w) {
  f(100*(1+(w(0)-0.5)/10), 100*(1+(w(1)-0.5)/10), 1000*(1+(w(2)-0.5)/10), 
    0.5*(1+(w(3)-0.5)/10))
}

draw 3000 from X
\end{verbatim}

This script produces a density plot of the sensitivity of the concentrations (when shift-clicking the Play button to run uninterrupted). For the sensitivity of the standard deviation of the concentrations, replace the result of \verb|f| with
\begin{verbatim}
  yield [observe(sqrt(var(a)),E), observe(sqrt(var(b)),E), 
         observe(sqrt(var(c)),E)]
\end{verbatim}
and run the script with LNA enabled.

\section{Data for Gibson Assembly}
\noindent
We report the experimental data employed for the Gibson assembly protocol in Section \ref{sec:ExpGIbson} (from \cite{gibson2009enzymatic} Figure 2a). For simplicity we report the output values of only species $O$ and omit unit of measurement, which are mM for concentration, $\mu$L for volume, Celsius degrees for temperature, and seconds for equilibration time:
\begin{align*}
   & d_1=((1, 0,0,0,1,20),(1,0),0)\quad d_2=((1, 1, 0,0,0,1,20),(1,120),0)\\
   &d_3=((1,  0,0,0,1,20),(1,240),0.05) \quad d_4=((1,  0,0,0,1,20),(1,360),0.56) \\ 
   &d_5=(( 1, 0,0,0,1,20),(1,480),0.8) \quad d_6=(( 1, 0,0,0,1,20),(1,660),0.86)\\ 
   &d_5=(( 1, 0,0,0,1,20),(1,840),0.9) \quad d_6=(( 1, 0,0,0,1,20),(1,960),0.88),
\end{align*}
where, for example, in $d_1$ we have that 
the initial concentration for $AB$ and $BA$ is 1 (note that in this case the optimization variables are $x_{BA}$ and $T$, hence in $d_1$ the vector $(1,0)$ represents the value assigned to those variables during the particular experiment) and all other species are not present at time $0$, volume and temperature at which the experiment is performed are $1$ and $20$ and the observed value for $O$ at the end of the protocol for $T=0$ is $0.$ We assume an additive Gaussian observation noise (noise in the collection of the data) with standard deviation $\sigma=0.1$.

\end{paracol}   


\reftitle{References}


\externalbibliography{yes}
\bibliography{references}

\begin{thebibliography}{999}

\bibitem[Murphy \em{et~al.}(2018)Murphy, Petersen, Phillips, Yordanov, and
  Dalchau]{murphy2018synthesizing}
Murphy, N.; Petersen, R.; Phillips, A.; Yordanov, B.; Dalchau, N.
\newblock Synthesizing and tuning stochastic chemical reaction networks with
  specified behaviours.
\newblock {\em Journal of The Royal Society Interface} {\bf 2018}, {\em
  15},~20180283.

\bibitem[Ananthanarayanan and Thies(2010)]{ananthanarayanan2010biocoder}
Ananthanarayanan, V.; Thies, W.
\newblock Biocoder: A programming language for standardizing and automating
  biology protocols.
\newblock {\em Journal of biological engineering} {\bf 2010}, {\em 4},~1--13.

\bibitem[Cardelli \em{et~al.}(2017)Cardelli, {\v{C}}e{\v{s}}ka, Fr{\"a}nzle,
  Kwiatkowska, Laurenti, Paoletti, and Whitby]{cardelli2017syntax}
Cardelli, L.; {\v{C}}e{\v{s}}ka, M.; Fr{\"a}nzle, M.; Kwiatkowska, M.;
  Laurenti, L.; Paoletti, N.; Whitby, M.
\newblock Syntax-guided optimal synthesis for chemical reaction networks.
\newblock  International Conference on Computer Aided Verification. Springer,
  2017, pp. 375--395.

\bibitem[Ang \em{et~al.}(2013)Ang, Harris, Hussey, Kil, and
  McMillen]{ang2013tuning}
Ang, J.; Harris, E.; Hussey, B.J.; Kil, R.; McMillen, D.R.
\newblock Tuning response curves for synthetic biology.
\newblock {\em ACS synthetic biology} {\bf 2013}, {\em 2},~547--567.

\bibitem[Abate \em{et~al.}(2018)Abate, Cardelli, Kwiatkowska, Laurenti, and
  Yordanov]{abate2018experimental}
Abate, A.; Cardelli, L.; Kwiatkowska, M.; Laurenti, L.; Yordanov, B.
\newblock Experimental biological protocols with formal semantics.
\newblock  International Conference on Computational Methods in Systems
  Biology. Springer,  2018, pp. 165--182.

\bibitem[Rasmussen \em{et~al.}(2006)Rasmussen, Williams, and
  Bach]{rasmussen2003gaussian}
Rasmussen, C.E.; Williams, C.K.; Bach, F.
\newblock {\em Gaussian Processes for Machine Learning}; MIT Press,  2006.

\bibitem[Van~Kampen(1992)]{Kampen1992b}
Van~Kampen, N.G.
\newblock {\em Stochastic processes in physics and chemistry}; Vol.~1,
  Elsevier,  1992.

\bibitem[Cardelli \em{et~al.}(2016)Cardelli, Kwiatkowska, and
  Laurenti]{cardelli2016stochastic}
Cardelli, L.; Kwiatkowska, M.; Laurenti, L.
\newblock Stochastic analysis of chemical reaction networks using linear noise
  approximation.
\newblock {\em Biosystems} {\bf 2016}, {\em 149},~26--33.

\bibitem[Gibson \em{et~al.}(2009)Gibson, Young, Chuang, Venter, Hutchison, and
  Smith]{gibson2009enzymatic}
Gibson, D.G.; Young, L.; Chuang, R.Y.; Venter, J.C.; Hutchison, C.A.; Smith,
  H.O.
\newblock Enzymatic assembly of DNA molecules up to several hundred kilobases.
\newblock {\em Nature methods} {\bf 2009}, {\em 6},~343--345.

\bibitem[Begley and Ellis(2012)]{begley2012raise}
Begley, C.G.; Ellis, L.M.
\newblock Raise standards for preclinical cancer research.
\newblock {\em Nature} {\bf 2012}, {\em 483},~531--533.

\bibitem[Ott \em{et~al.}(2018)Ott, Loveless, Curtis, Lesani, and
  Brisk]{ott2018bioscript}
Ott, J.; Loveless, T.; Curtis, C.; Lesani, M.; Brisk, P.
\newblock Bioscript: programming safe chemistry on laboratories-on-a-chip.
\newblock {\em Proceedings of the ACM on Programming Languages} {\bf 2018},
  {\em 2},~1--31.

\bibitem[Baker(2016)]{baker20161}
Baker, M.
\newblock 1,500 scientists lift the lid on reproducibility.
\newblock {\em Nature News} {\bf 2016}, {\em 533},~452.

\bibitem[Transcriptic()]{autoprot}
Transcriptic.
\newblock Autoprotocol.

\bibitem[Synthace()]{antha}
Synthace.
\newblock Antha.

\bibitem[Cardelli(2020)]{cardelli2020kaemika}
Cardelli, L.
\newblock Kaemika app: Integrating protocols and chemical simulation.
\newblock  International Conference on Computational Methods in Systems
  Biology. Springer,  2020, pp. 373--379.

\bibitem[Scott and Strachey(1971)]{scott1971toward}
Scott, D.; Strachey, C.
\newblock {\em Toward a mathematical semantics for computer languages}; Vol.~1,
  Oxford University Computing Laboratory, Programming Research Group Oxford,
  1971.

\bibitem[Cardelli(2013)]{Cardelli2010}
Cardelli, L.
\newblock Two-domain {DNA} strand displacement.
\newblock {\em Mathematical Structures in Computer Science} {\bf 2013}, {\em
  23},~247--271.

\bibitem[Bortolussi \em{et~al.}(2019)Bortolussi, Cardelli, Kwiatkowska, and
  Laurenti]{bortolussi2019central}
Bortolussi, L.; Cardelli, L.; Kwiatkowska, M.; Laurenti, L.
\newblock Central limit model checking.
\newblock {\em ACM Transactions on Computational Logic (TOCL)} {\bf 2019}, {\em
  20},~1--35.

\bibitem[Laurenti \em{et~al.}(2018)Laurenti, Csikasz-Nagy, Kwiatkowska, and
  Cardelli]{laurenti2018molecular}
Laurenti, L.; Csikasz-Nagy, A.; Kwiatkowska, M.; Cardelli, L.
\newblock Molecular Filters for Noise Reduction.
\newblock {\em Biophysical Journal} {\bf 2018}, {\em 114},~3000--3011.

\bibitem[Micchelli \em{et~al.}(2006)Micchelli, Xu, and
  Zhang]{micchelli2006universal}
Micchelli, C.A.; Xu, Y.; Zhang, H.
\newblock Universal Kernels.
\newblock {\em Journal of Machine Learning Research} {\bf 2006}, {\em 7}.

\bibitem[Boyd \em{et~al.}(2004)Boyd, Boyd, and Vandenberghe]{boyd2004convex}
Boyd, S.; Boyd, S.P.; Vandenberghe, L.
\newblock {\em Convex optimization}; Cambridge university press,  2004.

\bibitem[Newman \em{et~al.}(2019)Newman, Stephenson, Willsey, Nguyen,
  Takahashi, Strauss, and Ceze]{newman2019high}
Newman, S.; Stephenson, A.P.; Willsey, M.; Nguyen, B.H.; Takahashi, C.N.;
  Strauss, K.; Ceze, L.
\newblock High density DNA data storage library via dehydration with digital
  microfluidic retrieval.
\newblock {\em Nature communications} {\bf 2019}, {\em 10},~1--6.

\bibitem[Ethier and Kurtz(2009)]{ethier2009markov}
Ethier, S.N.; Kurtz, T.G.
\newblock {\em Markov processes: characterization and convergence}; Vol. 282,
  John Wiley \& Sons,  2009.

\bibitem[Schwabe \em{et~al.}(2012)Schwabe, Rybakova, and
  Bruggeman]{schwabe2012transcription}
Schwabe, A.; Rybakova, K.N.; Bruggeman, F.J.
\newblock Transcription stochasticity of complex gene regulation models.
\newblock {\em Biophysical Journal} {\bf 2012}, {\em 103},~1152--1161.

\bibitem[Leake(2014)]{leake2014analytical}
Leake, M.
\newblock Analytical tools for single-molecule fluorescence imaging in cellulo.
\newblock {\em Physical Chemistry Chemical Physics} {\bf 2014}, {\em
  16},~12635--12647.

\end{thebibliography}

\end{document}